\documentclass{article}

\usepackage{arxiv}

\usepackage[utf8]{inputenc} 
\usepackage[T1]{fontenc}    
\usepackage{cite}
\usepackage{hyperref}       
\hypersetup{hidelinks}
\usepackage{url}            
\usepackage{booktabs}       
\usepackage{amsfonts}       
\usepackage{nicefrac}       
\usepackage{microtype}      
\usepackage{graphicx}
\usepackage{natbib}
\usepackage{doi}
\usepackage{subcaption}

\title{Detection of LLM-Generated Java Code Using Discretized Nested Bigrams}

\author{ \href{https://orcid.org/0009-0003-0430-4407}{\includegraphics[scale=0.06]{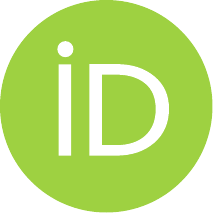}\hspace{1mm}Timothy Paek}\\
	College of Engineering and Computer Science\\
	Syracuse University\\
	Syracuse, NY 13210 \\
	\texttt{tipaek@syr.edu} \\
	\And
	\href{https://orcid.org/0000-0002-6149-6930}{\includegraphics[scale=0.06]{orcid.pdf}\hspace{1mm}Chilukuri Mohan} \\
	College of Engineering and Computer Science\\
	Syracuse University\\
	Syracuse, NY 13210 \\
	\texttt{ckmohan@syr.edu} \\
}

\renewcommand{\shorttitle}{Detection of LLM-Generated Java Code}

\hypersetup{
pdftitle={Detection of LLM-Generated Java Code Using Discretized Nested Bigrams},
pdfsubject={q-bio.NC, q-bio.QM},
pdfauthor={Timothy Paek, Chilukuri Mohan},
pdfkeywords={LLM-Generated Code, Code Authorship Attribution, GPT Code Detection, Large Language Models (LLMs), Stylometry Features.},
}

\begin{document}
\maketitle

\begin{abstract}
Large Language Models (LLMs) 
  are currently 
  used extensively to 
  generate code by professionals and students, motivating the development of tools 
  to detect LLM-generated code for applications such as academic integrity and cybersecurity. We address this authorship attribution 
  problem as a 
  binary classification task along with feature identification and extraction.
     We propose new    \textit{Discretized Nested Bigram Frequency} features on source code groups of various sizes. 
     Compared to prior work, improvements are obtained  by representing sparse information in dense membership bins.
  Experimental evaluation 
   demonstrated that our approach significantly outperformed a commonly used GPT code-detection API and baseline features, with accuracy exceeding 96\% compared to 72\% and 79\% respectively in detecting GPT-rewritten Java code fragments for 976 files with GPT 3.5 and GPT 4 using 12 features. We also outperformed three prior works on code author identification in a 40-author dataset.  Our approach scales well to larger data sets, and we achieved 99\% accuracy and 0.999 AUC for 76,089 files and over 1,000 authors with GPT 4o using 227 features.
\end{abstract}








\section{Introduction}
Detecting the authors of textual materials 
is a problem known as authorship attribution, 
and may be addressed using stylometric  techniques that 
perform quantitative analyses of authors' writing styles.
In addition to analyzing works of literature, the stylometric approach has been explored for the identification of authors of Python code, e.g., \citep{H18}.

The recent extensive use of large language models (LLMs) to generate code 
has raised serious concerns, especially in
academic environments where students 
claim GPT-generated code as their own.
Code generated by the same LLM may vary stylistically between samples, due to the multiplicity of sources used for training the LLM.
This motivates our research on source code 
authorship attribution. 
Applications include academic integrity concerns, software development process evaluation, and assessments of intellectual property ownership.


We propose a new approach to address this problem, splitting Java source code into code fragments 
before feature extraction, and then using binary classification to identify whether a code fragment 
varies substantially from other code fragments written by an author, so that
the model can distinguish between code written by different authors in the same file. 

The focus of our research  has been on creating features suited to this task.
We  have also created and tested our approach on datasets with combinations of LLMs,  formatted and non-formatted code. 
We 
assess
human-authored Java code rewritten by the LLM; this is a scalable approach to large dataset creation while simulating potential prompts.
Problem difficulty is increased since the LLM rewrites 
human-authored code 
written using
various coding styles. 

We treat LLM source code detection as a binary classification problem, and 
 achieved accuracy exceeding 96\%  in detecting Java code written by GPT 3.5 and GPT 4, 
     using a newly created 
    dataset of 976 files, 
   significantly outperforming a commonly used GPT code-detection API. 
   We also achieved accuracy exceeding 98\% 
   on another dataset with 40 authors and 3,021 files. Finally, we achieved 99\% accuracy in detecting GPT 4o with 76,089 files, demonstrating our approach's scalability. 
   New features (Discretized CodeBERT-embedded Nested Bigrams combined with Nested Bigram Frequencies) improved performance further, to 97.5\% and 99\% accuracy on the 
   two datasets, respectively. 
   
Section 2 discusses background including related work.
Section 3 presents our approach, and experimental results are provided in Section 4.
Section 5 summarizes and discusses limitations.

\section{Background}
This section summarizes  related work, the stylometric approach, 
and the primary features used in our research.

\subsection{Related Work}

In prior work \citep{Y17, O21, C22, H22, A23}, code authorship attribution has typically been addressed as a multi-class classification problem, 
attributing an entire source code file to a single author. However, this has the following limitations:
\\\\ 
1. Files 
    often have multiple authors, potentially resulting in model confusion. 
\\    
2. An LLM may have been used to generate only some snippets of code, which may be undetected because the rest of the code was written by the claimant author(s).
\\ 3. There is often limited data on 
    code from an individual, resulting in 
    unavailability of 
    sufficient data for effective training.\\ 

Additionally, current LLM detectors are trained on 
both code and non-code text, resulting in inadequate performance on code alone. 
Most utilized features are based on 
natural language processing, 
not addressing syntactic features specific to code, such as Abstract Syntax Tree (AST) features. 
Three such recent works are discussed below.


Oedingen et al. (2024)
address GPT code detection 
for Python code by utilizing Term Frequency - Inverse Term Frequency (TF-IDF), code2Vec, and human-generated features, 
achieving up to 0.98 accuracy on non-formatted code and 0.94 accuracy on formatted code. Their work 
tests 
only for code generated by GPT 3.5 (not GPT 4 or GPT 4o)
 based on manually provided 
 prompts.

Ye et al. (2024) also 
follow a similar approach, rewriting both human-authored and machine code, and testing various similarity metrics, 
achieving up to 0.8325 and 0.8623 accuracy on APS and MBPP benchmarks for Python, respectively, using GPT 3.5 to rewrite the code; they achieved 0.9082 accuracy in detecting C++ code written by GPT 3.5. 

Xu and Sheng (2024) use masked perturbation and a fine-tuned CodeBERT model on datasets created from CodeNET and AIGCode programming problems in various languages, using text-davinci-003 to generate solutions to problem descriptions or translate code into another language. However, this approach only achieved 0.82 AUC for detecting LLM-authored Java code, 
substantially lower than the results from our approach.

\subsection{Stylometry Features}

Stylometry involves describing 
authors' unique coding styles using quantifiable features 
such as: 
\\\\ 
1. Lexical features, focusing on 
specific tokens in the code, e.g., 
keywords, identifiers, and length.
\\ 
2. syntactic features, addressing the structure of the code, e.g.,  
n-grams and control structures.
\\ 3. 
Layout features, relating to code 
formatting, 
e.g., white-space, line-breaks, and indentation.\\

Among these categories, all useful in distinguishing between author coding styles, syntactic features tend to provide the most information specific to certain code authors \citep{C15}.  
Emphasizing syntactic features makes the model 
more robust to layout obfuscation,
e.g., when those who use LLMs (to generate code)  
reformat the code, changing the amount of white-space.
%

N-gram Abstract Syntax Tree (AST) features have been shown to be powerful in capturing the syntax of source code. 
In particular, \textit{Nested Bigrams (NB)} \citep{H20} 
increase the specificity of features by 
focusing on a sub-tree of the AST, instead of just a node. 

\section{Methods}
This section describes the dataset generation process, followed by the feature extraction approach and the machine learning models used.

\subsection{Dataset Generation}

To the best of our knowledge,  no publicly available datasets exist for LLM-generated Java code classification. 
Also, 
 different LLMs may correspond to different  writing styles, and 
variations in the prompts may also result in variations of writing style. 
Hence we created two new datasets: (1) \textit{GPT Dataset}, a smaller one focusing on GPT 3.5 and GPT 4 models; (2) \textit{GPT GCJ Dataset}, a larger one focusing on GPT 4o. We have made both 
datasets publicly available on GitHub \citep{P24_Java, P24_GCJ}. 
\\ \\
The following process was used to obtain the GPT Dataset:
\begin{enumerate}
    \item 666 Java source code files from 11 different authors' GitHub pages were acquired from another 
    dataset \citep{Y17}.
    \item 5 of the 11 authors' files were passed through either ChatGPT 3.5 or Bing GPT 4 in a rewriting task,
using the prompt: 
"The messages I send you will be in Java code. I want you to rewrite all of it while maintaining functionality."
    \item The entire 
    file was passed through BingGPT (4000 character limit) and ChatGPT without additional prompting; the resulting code was then pasted into a new file.
    \item The resulting files were either saved without additional formatting or were formatted using VSCode's format. 
\end{enumerate}
This resulted in 976  files, including 
   666 files of original authors,
     108 files rewritten  using Bing GPT 4 (61 formatted, 47 non-formatted); and
     202 files rewritten  using ChatGPT 3.5 (59 formatted, 143 non-formatted).
\\ \\    
The following process was used to obtain the GPT GCJ Dataset:

\begin{enumerate}
    \item 58,524 human-authored Java source code files from over 1,000 participants were retrieved from the 2020 Google Code Jam competition.
    \item 17,565 of these files were rewritten by GPT 4o API with the prompt: "This is java code. Rewrite it entirely while maintaining functionality."
\end{enumerate}
This resulted in 76,089 files, including
  58,524 files of original authors, 
 and 17,565 files rewritten using GPT 4o API.
    
By having the LLM rewrite existing code, we attempt to ensure that the LLM code is similar to the original, 
to make it 
difficult to distinguish between the classes. The rewriting task intends to simulate different potential input prompts in a scalable manner. 
The rewritten code typically has rewritten variable names and slightly different formatting while still achieving the same code functionality. Additionally, some rewritten code was reformatted using VSCode to emulate what may occur in real-world scenarios for the GPT Dataset. 

Code was broken up into chunks called \textit{code groups}, whose size (number of lines) 
varied from 10-70, yielding different datasets.
For discretized features, an additional parameter, \textit{bin width}, determines how many features are allocated to a single bin.


\subsection{Feature Extraction}
For non-syntactic features, we use mean line length, mean comment length, and the numbers of spaces (Whitespace), statement words (e.g., "if", "for", etc.), tabs, underscores, and empty lines.
These are normalized by dividing by the length of the code group in characters; this captures some 
contextual information. 

The rest of this section focuses on syntactic features.
We begin by observing that in nested bigrams (NB), nodes 
include attribute information such as the names of input variables, meaning that if another node does not contain the exact attribute information, it will not match. 
Feature specificity further increases  
due to the increased sub-tree relationships captured by nested bigrams. 
However, their sparsity (small number of occurences in the dataset) 
appeared to diminish 
performance. 
Using Principal Component Analysis 
and Autoencoding 
did not help, motivating the exploration of new features.

Increased feature specificity usually implies that the number of features 
increases. For example, a dataset with 1,000 files split into code groups of size 10 can have 90,000+ features from NB alone. 
However, using such highly specific features results in 
increasing the storage requirements for 
the feature information.
We address these concerns by exploring new features that use \textit{equal width discretization }(combining multiple features into a single bin) 
and 
\textit{Compressed} Nested Bigrams (CNB) which 
exclude attribute information, thus reducing the specificity of the feature. 
Discretization  
results in features that represent soft logic membership for subsets of each author's code data.

For syntactic features, we tested the following AST features separately (making different datasets for each); we also experimented with transformer embeddings and frequencies for these features:
\begin{itemize}

\item 
CodeBERT-embedded Nodes 
\textbf{(CBN)} use CodeBERT \citep{R21} to extract CLS token means for AST nodes.
 \item
Nested Bigram Frequencies \textbf{(NB-F)}   use subtrees of an AST.
\item 
Compressed Nested Bigrams Frequencies 
\textbf{(CNB-F)}  exclude attribute information, 
reducing variation in the number of  features, thus resulting in  a smaller and less sparse dataset, at the potential cost of information loss. 
\item 
Equal Width Discretized Nested Bigram Frequencies 
\textbf{(EWD-NB-F)} are discretized nested bigrams, where each bin is a simple summation of the frequency of nested bigrams within the range for the bin. 
\item 
Equal Width Discretized CodeBERT-Embedded Nested Bigram CLS Means 
\textbf{(EWD-CBNB-CM)} are similar 
but contain additional information from the transformer CLS tokens providing  contextual information. 


\end{itemize}




\begin{figure*}
   \includegraphics[width=\textwidth]{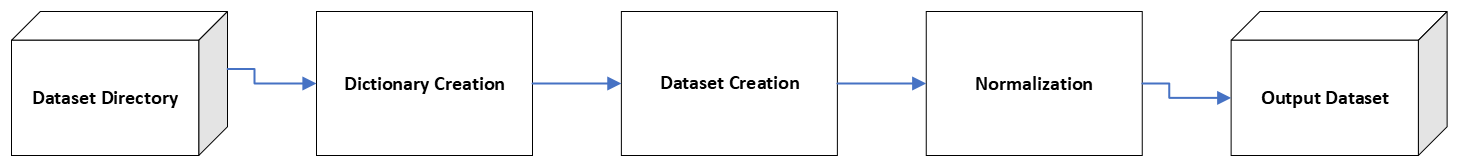}
    \caption{Feature Extraction 
Flowchart}
 \label{fig-main}
\end{figure*}

\begin{figure}[h]
\centering
\begin{subfigure}[b]{0.45\textwidth} 
     \includegraphics[width=\textwidth]{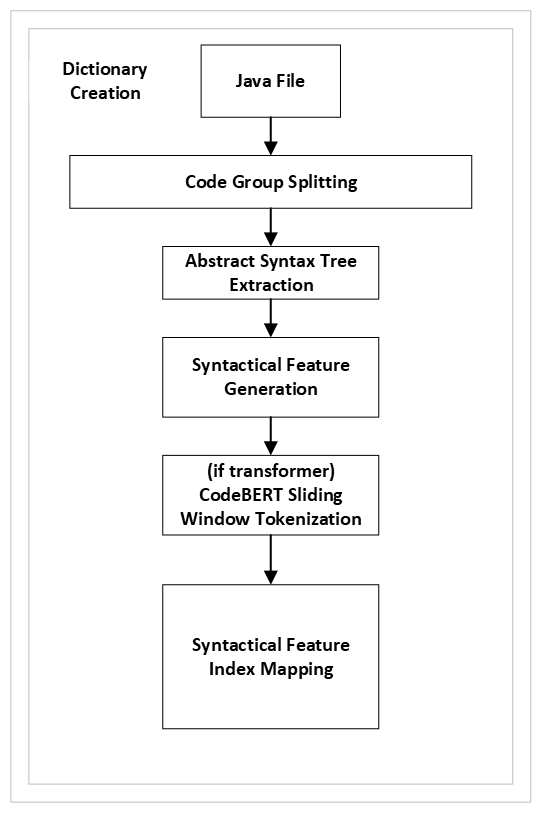}
    \caption{Dictionary Creation}
    \label{fig-dictionary}
\end{subfigure}
\begin{subfigure}[b]{0.45\textwidth}

    \includegraphics[width=\textwidth]{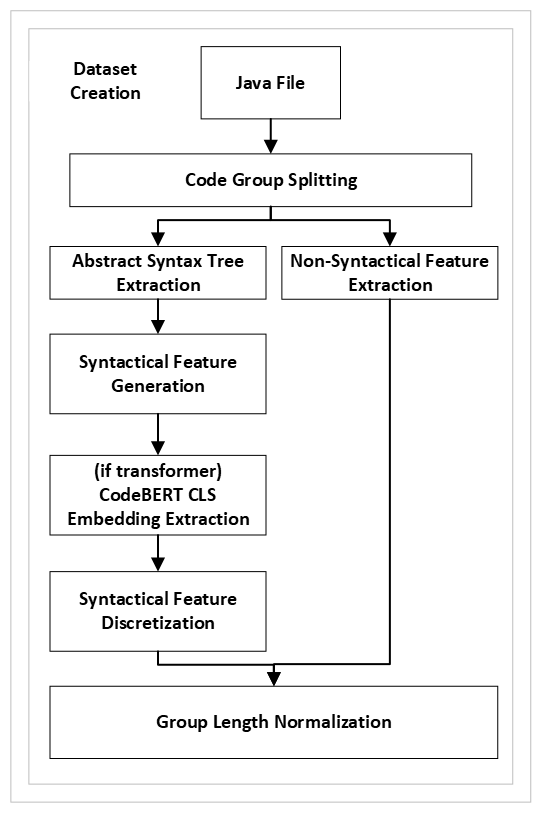}%
    \caption{Dataset Creation}
      \label{fig-dataset}
\end{subfigure}       
\caption{Two major components of the Feature Extraction process}
\label{fig-dictionary-dataset}
\end{figure}

Our feature extraction process is illustrated in Figure \ref{fig-main}. We take an input dataset and pass it through a dictionary creation block where features are mapped to temporary indices for calculating dataframe columns as shown in  Figure \ref{fig-dictionary}. We then utilize the dictionary in a dataset creation block where feature values are added to appropriate locations as documented by Figure \ref{fig-dataset}, and normalize the data. 
\begin{enumerate} 
\item 
For frequency-based features such as NB-F, CNB-F, and EWD-NB-F, we first extract 
[compressed] nested bigrams from the file according to the line bounds for the current code group. 
\item 
For transformer-based features, 
we first extract nested bigrams from the file according to the line bounds of the current group, then 
apply the CodeBERT tokenizer. We use 
a sliding window of length 512 tokens (the maximum input for CodeBERT), adding its string representation to the dictionary. 
We also tested CLS maximum and minimum on nested bigrams,
but these did not perform as well as the mean.

\item 
For non-discretized features, such as NB-F and CNB-F, we add the count of that feature in the code group to the index returned by the dictionary for that feature.
\item 
For discretized features such as EWD-NB-F and EWD-CBNB-CM, we again use the dictionary to calculate the dataset column index for each NB or CNB as follows: 
\begin{equation}
    \label{eqn:bin_index_equation}
        \frac{(i - s_1)} {b} + s_2
\end{equation}
where $i$ is the dictionary value associated with the feature; $b$ is the bin width; and $s_1$ is the start position of binning within the dictionary and $s_2$ is the start position of the binning within the dataframe. The first 10 values are reserved for non-syntactic features in both the dictionary and dataset. For datasets with only one discretized syntactic feature, $s_1$ and $s_2$ are $10$. For more than one discretized feature, $s_1$ and $s_2$ vary. Therefore, each bin has the value:
\begin{equation}
    \label{eqn:ewd_bin_equation}
        \sum_{i=a}^{a+w} 
        \frac{F_i}{c}
\end{equation}
where $a$ is the bin's starting index, 
$w$ is the width of the bin, $F_i$ 
could either be a frequency or the mean of the CLS token, and $c$ is the number of characters in the code group. 
\end{enumerate}

\subsection{Machine Learning Models Used} 

Our focus in this research has been on identifying the best features to enable identification of LLM-generated code using existing ML models or algorithms, not on improving the latter.
For many problems, ensembles of machine learning models have been shown to be 
more effective than individual models  \citep{K20}. 
This motivates us to use 
Random Forests \citep{B01}, XGBoost \citep{C16}, Light Gradient Boosting Machine (LGBM) \citep{K17}, and CatBoost \citep{P18} ensemble approaches,  whose code was imported from various publicly available sources, i.e.,  \textit{sklearn.ensemble, xgboost, lightgbm}, and \textit{catboost. }

Ensemble execution is deterministic; 
all random states for these models were initialized to the same value (42) for repeatability, with all other hyperparameters set to constant values. Experiments with other initial states showed little difference. 
The following values were used for the parameters of these machine learning approaches:
\begin{itemize}
    \item \textbf{Random Forest:}
    \begin{itemize}
    \item n\_estimators = 100
    \item criterion = gini
    \item min\_samples\_split = 2
    \item min\_samples\_leaf = 1
    \end{itemize}
\item \textbf{XGBoost:}
\begin{itemize}
    \item booster = gbtree
    \item eta = 0.3
    \item max\_depth = 6
    \item sampling\_method = uniform
    \item grow\_policy = depthwise
\end{itemize}
\item \textbf{LGBM:}
\begin{itemize}
    \item objective = regression
    \item boosting = gbdt
    \item data\_sample\_strategy = bagging
    \item num\_iterations = 100
    \item num\_leaves = 31
    \item tree\_learner = serial
\end{itemize}
\item \textbf{CatBoost:}
\begin{itemize}
    \item iterations = 1000
    \item learning\_rate = 0.03
    \item l2\_leaf\_reg = 3.0
    \item depth = 6
    \item max\_leaves = 31
    \item fold\_len\_multiplier = 2
\end{itemize}
\end{itemize}

We achieved satisfactory results 
using default parameter values. 
Performance was sufficiently high, so that additional tuning was not needed. 
Results were evaluated using
well-known measures: 
accuracy, F1 score, 
 Area under the ROC Curve (AUC), and Precision.


\section{Experimental Results}
We 
evaluated our approach on 
the newly generated GPT and GPT GCJ datasets (described earlier)
as well as 
a 40-author Java Dataset with 3,021 files \citep{Y17} with 15 authors considered the positive class,
splitting training and testing sets. 
Dataset features are subjected to Winsorized normalization;
extreme values are mapped to 0 or 1, and 
 each non-extremal $x$ is mapped to $(x-x_{5\%})/(x_{95\%}-x_{5\%})$, where $x_{k\%}$ refers to the $k$th percentile.
%
%
%

We compared our results with ZeroGPT, an online service (API) that lets users detect whether input text was generated using a GPT model. We tested the API on the same code groups and classified the model's output by checking if
its probability score (of being generated by GPT) $>$0.5. 

\subsection{GPT Dataset Detection Results}

We compare the performance of group sizes 10-70 for various syntactic features, 
using Random Forest, XGBoost, LGBM, and CatBoost. 
Results shown are averages 
over all the ensembles tested.  The bin width is optimized for accuracy for discretized features, though 
the bin width 
did not affect results, as long as it is sufficiently large. Table \ref{GPT-results} shows only the results for group sizes 10, 40, and 70, omitting 20, 30, 50, and 60 for space reasons. 


Results are summarized below with average values across all code groups:

\begin{table*}[t]
\centering
\caption{GPT Detection Results}
\begin{tabular}{||c||c|c|c|c|c||}
\hline 
\textbf{Dataset} & \textbf{Group Size} & \textbf{Accuracy} & \textbf{F1} & \textbf{AUC} & \textbf{Precision} \\
\hline 
API & 10 & 0.727 & 0.16 & 0.5 & 0.16 \\
& 40 & 0.72 & 0.13 & 0.48 & 0.15 \\
& 70 & 0.73 & 0.13 & 0.48 & 0.18 \\
\hline 
CBN& 20 & 0.79 & 0.2 & 0.82 & 0.32 \\
& 40 & 0.76 & 0.17 & 0.78 & 0.35 \\
& 70 & 0.77 & 0.19 & 0.76 & 0.35 \\
\hline
NB-F & 10 & 0.848 & 0.4 & 0.85 & 0.63 \\
& 40 & 0.79 & 0.28 & 0.85 & 0.52 \\
& 70 & 0.79 & 0.31 & 0.83 & 0.59 \\
\hline 
CNB-F & 10 & 0.86 & 0.48 & 0.87 & 0.65 \\
& 40 & 0.85 & 0.57 & 0.9 & 0.73 \\
& 70 & 0.84 & 0.54 & 0.9 & 0.76 \\
\hline 
EWD-CBNB-CM & 10 & 0.95 & 0.67 & 0.96 & 0.69 \\
& 40 & 0.96 & 0.82 & 0.98 & 0.84 \\
& 70 & 0.96 & 0.81 & 0.98 & 0.85 \\
 \hline 
EWD-NB-F & 10 & 0.95 & 0.67 & 0.96 & 0.7 \\
& 40 & 0.97 & 0.83 & 0.97 & 0.86 \\
& 70 & 0.96 & 0.79 & 0.98 & 0.84 \\
\hline 
\textbf{EWD-NB-F + EWD-CBNB-CM} & 10 & 0.97 & 0.71 & 0.97 & 0.76 \\
& 40 & 0.98 & 0.79 & 0.97 & 0.85 \\
& 70 & 0.97 & 0.8 & 0.98 & 0.84 
\\ \hline 
\end{tabular}
    
    \label{GPT-results}
\end{table*}

\textbf{Accuracy:} 
API:  0.72; CBN; 0.78; NB-F: 0.82; CNB-F: 0.86; EWD-CBNB-CM: 0.957; EWD-NB-F: 0.960; EWD-NB-F in combination with EWD-CBNB-CM: \textbf{0.974.}

\textbf{F1:} 
API: 0.14; CBN: 0.17; NB-F: 0.33; CNB-F: 0.56; EWD-CBNB-CM: 0.767; EWD-NB-F: 0.774; EWD-NB-F in combination with EWD-CBNB-CM: \textbf{0.781.}

\textbf{AUC: }
API: 0.49; CBN: 0.78; NB-F: 0.84; CNB-F: 0.899; EWD-CBNB-CM: 0.974; EWD-NB-F: 0.972; EWD-NB-F in combination with EWD-CBNB-CM:\textbf{ 0.979.}

\textbf{Precision:} 
API: 0.16; CBN: 0.33; NB-F: 0.58; CNB-F: 0.74; EWD-CBNB-CM: 0.812; EWD-NB-F: 0.826; EWD-NB-F in combination with EWD-CBNB-CM: \textbf{0.838.}

The performance of ensembles for the same dataset was similar for XGBoost, LGBM, and CatBoost for all metrics tested. Random Forest performs slightly worse.  
Performance was only marginally 
affected by group size.

We 
achieved high performance in accuracy, F1, AUC, and precision scores in detecting GPT in the Java GPT Dataset. Even the worst performing syntactic feature NB-F significantly outperformed the ZeroGPT API in all code group sizes and in all metrics tested. As the API was primarily meant to be used in detecting GPT generated text, it failed to perform well in detecting GPT generated code, while our approach performed much better. 

CNB-F performed better than NB-F in all cases, 
since it has significantly fewer features. As an example, the NB-F dataset for group size 40 contains 13.5 thousand features, whereas CNB-F contains 52 features for all group sizes. 

EWD-CBNB-CM and EWD-NB-F perform significantly better than CNB-F, 
due to reducing the sparsity of the dataset and the density of the information. By representing all the information present in NB-F in dense bins, they retain diverse information 
with a small dimensionality, 
just 12 features.

Although EWD-CBNB-CM in combination with EWD-NB-F performs about 1\% better than EWD-NB-F, 
generating transformer embeddings for large quantities of Nested Bigram sliding window inputs requires significantly more computation than NB-F alone. 
Thus, EWD-NB-F alone is perhaps the best approach. 

We also performed t-tests for all pairwise combinations of group size 30 feature datasets with respect to accuracy, the same way it was calculated for the Table \ref{GPT-results},
using random initialization and random 
training vs. testing data splitting. 
The results for all pairs of datasets had a combined p-value $<10^{-3}$, 
implying that the differences in performances between datasets are 
not attributable to random variations. 
Similar conclusions are drawn from Table \ref{GPT-misc-metrics}. 
Deviation in performance is also found to be very low at 0.008 standard deviation at most, with discretized features having up to 0.003 in standard deviation. 

\begin{table*}[t]
\centering
\caption{GPT Group Size 30 Associated Accuracy Metrics}
\begin{tabular}{||c||c|c|c|c|c|}
\hline
\textbf{Dataset} & \textbf{Mean} & \textbf{Median} & \textbf{Std Dev} & \textbf{10th Pctl} & \textbf{90th Pctl} \\
\hline
NB-F & 0.821 & 0.821 & 0.008 & 0.808 & 0.831 \\
CNB-F & 0.876 & 0.875 & 0.007 & 0.868 & 0.886 \\
EWD-CBNB-CM & 0.963 & 0.963 & 0.003 & 0.959 & 0.966 \\
EWD-NB-F & 0.962 & 0.962 & 0.003 & 0.958 & 0.965 \\
EWD-NB-F + EWD-CBNB-CM & 0.974 & 0.974 & 0.002 & 0.971 & 0.977 \\
\hline
\end{tabular}
    
    \label{GPT-misc-metrics}
\end{table*}

\subsection{40-Author Anomaly Detection Results}



Similar comparisons were carried out with the 40-author data set \citep{Y17}.
Table \ref{40-results} shows only the results for group sizes 10, 40, and 70, again omitting 20, 30, 50, and 60 for space reasons. Results are summarized below with average values across all code groups:

\begin{table*}
\caption{Results for 40 Author Dataset}
\centering{
\begin{tabular}{||c||c|c|c|c|c||}
\hline
\textbf{Dataset} & \textbf{Group Size} & \textbf{Accuracy} & \textbf{F1} & \textbf{AUC} & \textbf{Precision} \\ \hline
NB-F & 10 & 0.94 & 0.73 & 0.96 & 0.85 \\
& 40 & 0.95 & 0.75 & 0.97 & 0.9 \\
& 70 & 0.95 & 0.77 & 0.98 & 0.95 \\
\hline 
CNB-F & 10 & 0.95 & 0.75 & 0.97 & 0.9 \\
& 40 & 0.96 & 0.8 & 0.98 & 0.92 \\
& 70 & 0.96 & 0.81 & 0.99 & 0.97 \\
\hline 
EWD-NB-F & 10 & 0.97 & 0.86 & 0.98 & 0.93 \\
& 40 & 0.98 & 0.91 & 0.99 & 0.95 \\
& 70 & 0.98 & 0.94 & 1 & 0.98 \\
\hline 
\textbf{EWD-CBNB-CM + EWD-NB-F} & 10 & 0.98 & 0.87 & 0.99 & 0.92 \\
& 40 & 0.99 & 0.92 & 0.99 & 0.96 \\
& 70 & 0.99 & 0.95 & 1 & 0.98 \\
\hline
\end{tabular}
}

\label{40-results}  
\end{table*}

\textbf{Accuracy:} 
NB-F: 0.95; CNB-F: 0.96; EWD-NB-F:0.98; EWD-NB-F in combination with EWD-CBNB-CM:\textbf{ 0.99}.

\textbf{F1:} 
NB-F: 0.75; CNB-F: 0.78; EWD-NB-F:0.91; EWD-NB-F in combination with EWD-CBNB-CM: \textbf{0.92}.

\textbf{AUC: }
NB-F:  0.97; CNB-F: 0.98; EWD-NB-F: 0.993; EWD-NB-F in combination with EWD-CBNB-CM: \textbf{0.996}. 

\textbf{Precision:} 
NB-F: 0.91; CNB-F: 0.93; EWD-NB-F in combination with EWD-CBNB-CM: 0.94; EWD-NB-F: \textbf{0.96}.

Results obtained with XGB, CatBoost, and LGBM, were almost identical; the Random Forest approach performed slightly worse.
Code group size did not significantly affect the performance in terms of accuracy, F1, and AUC, but affected precision.


We achieved high performance in all code groups using the same methods that we used for the GPT detection task. We also see the same trends in performance for the metrics we tested as measured by the maximum value across code groups; NB-F is outperformed by CNB-F; CNB-F is outperformed by EWD-NB-F; EWD-NB-F is outperformed by EWD-CBNB-CM. Performance for all features is higher in this dataset. Java GPT Dataset contains a subset of the authors in the 40-author dataset which implies that the GPT rewritten task is more difficult than typical binary code classification.
Results in Table \ref{40-comparison} compare our approach with that of three other recent works that address code author identification \citep{A23, O21, Y17}, demonstrating better performance.

\begin{table*}[t] 
\caption{Comparison of performance with different features, for 40 Author Dataset} 
\centering{
\begin{tabular}{||c|c|c|c|c|c|c|c||}
\hline
\textbf{Approach} & \textbf{Type} & \textbf{Group Size} & \textbf{Features} & \textbf{Accuracy} & \textbf{F1} \\ 
\hline
code2seq \citep{O21} & Multi-Class & Full File & -- & 0.967 & 0.9405 \\
Bigram \citep{A23} & Multi-Class & Full File & 1500 & 0.937 & -- \\
PSOBP \citep{Y17} & Multi-Class & Full File & -- & 0.91 & -- \\ 
\textbf{EWD-NB-F} & Binary & 70 & 61 & 0.98 & 0.94 \\
\textbf{EWD-CBNB-CM + EWD-NB-F} & Binary & 70 & 85 & 0.99 & 0.95 \\
\hline
\end{tabular}
}

\label{40-comparison}
\end{table*}

\begin{table*}[h!]
\caption{GPT GCJ Detection Results with EWD-NB-F} 
\centering{
\begin{tabular}{
||c||c|c|c|c|}
\hline
\textbf{Group Size} & \textbf{Accuracy} & \textbf{F1} & \textbf{AUC} & \textbf{Precision} \\
\hline
10 & 0.98 & 0.87 & 0.987 & 0.92 \\
 40 & 0.99 & 0.95 & 0.998 & 0.98 \\
 70 & 0.99 & 0.95 & 0.999 & 0.98 \\
\hline
\end{tabular}
}

  \label{GPT-GCJ-Results} 
\end{table*}

Comparing our results with other approaches applied to the same dataset, we predict binary classes on code groups while previous papers predict multiple classes on entire files. 
Despite this difference, our approach performed substantially better. We maintain a small dimensionality while improving  performance.  

\subsection{GPT GCJ Dataset Detection Results}

Finally, we 
addressed the GPT GCJ Dataset, which has up to 143,776 code groups for group size 10 and over 1,000 authors. We focus primarily on EWD-NB-F. Similar 
performance may 
be achievable with other features. 
However, we did not test on those features due to the high 
computational
requirements, e.g., 
creation of a single dataset for group size 30 for EWD-CBNB-CM 
required over 3 days.
We use bin width 3000, though any width $>$ 2000 performed nearly the same. Results are detailed in Table \ref{GPT-GCJ-Results}.

Performance remained 
high despite the significant scale-up in authors and files. Accuracy was maintained at 0.98-0.99; F1 score was 
0.92-0.95 (excluding group size 10); AUC was 
0.987-0.999; and precision was 
0.97-0.98 (excluding group size 10). Performance was similar for different ensemble models, with Random Forest models performing slightly better than others.
Additionally, performance  was stable across various experiments, e.g., with 40 random states for ensemble and train test split initialization, mean accuracy was 0.9855 with standard deviation  0.0003, for group size 30. 

Feature dimensionality remained low at 227-239 total features for bin width size 3,000. However, we find similar performance 
over a certain size. 

\subsection{Computational Effort Required}
We used an AMD Ryzen 9 5950X 16-Core Processor with 4.00 GHz, 64 GB RAM, Quadro RTX 5000. 
The feature extraction and dataset generation process required about 3 minutes for the GPT dataset and 10 minutes for the 40-author dataset.
For evaluating discretized GPT datasets using the ensemble models, about 1 minute was required (on average) to complete the experiment for GPT Dataset and 48 minutes for GPT GCJ Dataset, evaluating 7 datasets for each group size. For discretized 40-author datasets, the computational time required by the models was  3 minutes, 28 seconds (on average) 
for all 7 datasets.

\section{Conclusion}
This paper has addressed the problem of detecting LLM-generated fragments in Java code.
Source code files were split into code groups (10-70  lines each) and categorized using new stylometric features (EWD-NB-F, EWD-CBNB-CM, and CNB-F); the subsequent application of well-known machine learning models 
resulted in
very high accuracy.

Two datasets were created (and have been made available for other researchers) specific to Java GPT code detection: (1) GPT Dataset - with 15 human authors in addition to LLM (GPT 3.5 and GPT 4) rewritten code, (2) GPT GCJ Dataset - with 1k+ human authors in addition to GPT 4o rewritten code, where the LLM was prompted to rewrite the 
code while maintaining functionality. 
 Testing was performed using these as well as another  40-author dataset.


We achieved up to 0.977 accuracy on the GPT Dataset by combining EWD-NB-F and EWD-CBNB-CM 
using only 17 features, outperforming the commonly utilized GPTZero API which only achieves 0.73 accuracy. 
For the 40-author dataset, 
we achieved 0.99 accuracy with 
85 features.
We also achieved up to 0.99 accuracy on the GPT GCJ Dataset with 76k+ files using EWD-NB-F. 

Our approach 
significantly outperformed three other recent approaches. 
Consistent results were observed for different settings of code group sizes.  For example, 
we achieved 0.97-0.98 accuracy on the GPT dataset, whereas the ZeroGPT 
API results were substantially worse (accuracy 0.73).  When comparing feature sets, 
the improvement was substantial for one data set (from baseline feature set accuracy 0.79-0.86 to 0.95-0.98); 
even in the other (easier) task, the accuracy was improved from 0.94-0.95 (baseline feature set) to  0.98-0.99.

Future work includes exploring whether other mathematical operations in discretization could yield better results and making experiments more akin to real-world scenarios (code written by students and professionals). 
A real world LLM code detector would need to understand many LLM coding styles and be accurate even when given coding styles not seen in the input dataset; this needs to be assessed, and further improvements may be necessary to ensure such robustness. 
Finally, we note that the focus of this work was to find useful features for the task at hand, not on finding the best machine learning model or algorithm for the author identification task; 
performance improvements 
may be obtainable using other machine learning models (or algorithms). 
 


\subsection{Acknowledgements} The first author, an undergraduate student, gratefully acknowledges support from ICCAE
and the Renee Crown 
Honors program 
at Syracuse 
University and the Information Technology Services who provided a GPU in April 2024 for the final experiments whose results are reported.
\newpage
\bibliographystyle{unsrtnat}
\bibliography{references}
\end{document}